%% file: thin.tex
\documentclass{article}

\input{imports}
\begin{document}

\title{Builtin Types viewed as Inductive Families
}
\author{Guillaume Allais}

\maketitle

\togglefalse{BLIND}

\begin{abstract}
  \input{abstract}
\end{abstract}


\input{introduction}

\input{optimisation}

\input{QTT}
\input{CdB}

\input{efficient}

\input{conclusion}
\input{acknowledgements}

\bibliography{thin}

\appendix

\input{appendix-axioms}
\input{appendix-limitations}

\end{document}

%% file: imports.tex
\usepackage{etoolbox}
\usepackage{catchfilebetweentags}
\input{robust-catch}

\usepackage{idris2}

\usepackage[frozencache]{minted}
\usepackage{cleveref}

\usepackage{todonotes}
\setuptodonotes{inline}

\usepackage{tikz}
\usetikzlibrary{shapes.geometric}

\newcommand{\typos}{TypOS}
\newcommand{\idris}{Idris 2}
\newcommand{\coq}{Coq}
\newcommand{\agda}{Agda}
\newcommand{\bitcons}[2]{#1\!\cdot{}\!#2}

\newtoggle{BLIND}

%% file: abstract.tex
State of the art optimisation passes for dependently typed languages can help
erase the redundant information typical of invariant-rich data structures and
programs.
These automated processes do not dramatically change the \emph{structure} of
the data, even though more efficient representations could be available.

Using Quantitative Type Theory, we demonstrate how to define an invariant-rich,
typechecking time data structure packing an efficient runtime representation
together with runtime irrelevant invariants. The compiler can then aggressively
erase all such invariants during compilation.

Unlike other approaches, the complexity of the resulting representation is
entirely predictable, we do not require both representations to have the
same structure, and yet we are able to seamlessly program as if we were
using the high-level structure.

%% file: introduction.tex
\section{Introduction}

Dependently typed languages have empowered users to precisely describe their domain
of discourse by using inductive families~\cite{DBLP:journals/fac/Dybjer94}.
Programmers can bake crucial invariants directly into their definitions thus refining
both their functions' inputs and outputs.
The constrained inputs allow them to only consider the relevant cases during pattern
matching, while the refined outputs guarantee that client code can safely rely on the
invariants being maintained.
This programming style is dubbed `correct by construction'.

However, relying on inductive families can have a non-negligible runtime cost if
the host language is compiling them naïvely. And even state of the art optimisation
passes for dependently typed languages cannot make miracles: if the source code is
not efficient, the executable will not be either.

A state of the art compiler will for instance successfully compile length-indexed
lists to mere lists thus reducing the space complexity from quadratic to linear
in the size of the list.
But, confronted with a list of booleans whose length is statically known to be
less than 64, it will fail to pack it into a single machine word thus spending
linear space when constant would have sufficed.

In \cref{sec:optimisation-example}, we will look at an optimisation example
that highlights both the strengths and the limitations of the current state
of the art when it comes to removing the runtime overheads potentially
incurred by using inductive families.

In \cref{sec:quantitativeTT} we will give a quick introduction to Quantitative
Type Theory, the expressive language that grants programmers the ability
to have both strong invariants and, reliably, a very efficient runtime
representation.

In \cref{sec:codebruijn} we will look at an inductive family that
\iftoggle{BLIND}
  {the \typos{} project~\cite{MANUAL:talk/types/Allais22}
    uses in a performance-critical way}
  {we use in a performance-critical way in the
    \typos{} project~\cite{MANUAL:talk/types/Allais22}}
and whose compilation suffers from the limitations highlighted in
\cref{sec:optimisation-example}.
\iftoggle{BLIND}{The project's authors}{Our}
current and unsatisfactory approach is to rely on the safe and convenient
inductive family when experimenting in Agda and then replace it with an unsafe
but vastly more efficient representation in
\iftoggle{BLIND}{their}{our} actual Haskell implementation.

Finally in \cref{sec:efficient}, we will study the actual implementation of
our efficient and invariant-rich solution implemented in \idris{}. We will
also demonstrate that we can recover almost all the conveniences of programming
with inductive families thanks to smart constructors and views.

%% file: optimisation.tex
\section{An Optimisation Example}\label{sec:optimisation-example}

The prototypical examples of the naïve compilation of inductive families being
inefficient are probably the types of vectors (\IdrisType{Vect})
and finite numbers (\IdrisType{Fin}).
Their interplay is demonstrated by the \IdrisFunction{lookup} function.
Let us study this example and how successive optimisation passes can, in this
instance, get rid of the overhead introduced by using indexed families over
plain data.

A vector is a length-indexed list.
The type \IdrisType{Vect} is parameterised by the type of values it stores
and indexed over a natural number corresponding to its length.
More concretely, its \IdrisData{Nil} constructor builds an empty vector of size
\IdrisData{Z} (i.e. zero),
and its \IdrisData{(::)} (pronounced `cons') constructor combines a
value of type \IdrisBound{a} (the head) and a subvector of size \IdrisBound{n}
(the tail) to build a vector of size (\IdrisData{S} \IdrisBound{n})
(i.e. successor of n).

\ExecuteMetaData[Lookup.idr.tex]{vect}

The size \IdrisBound{n} is not explicitly bound in the type of \IdrisData{(::)}.
In \idris{}, this means that it is automatically generalised over in a prenex
manner reminiscent of the handling of free type variables in languages in the
ML family.
This makes it an implicit argument of the constructor. Consequently, given that
\IdrisType{Nat} is a type of \emph{unary} natural numbers, a naïve runtime
representation of a {(\IdrisType{Vect} \IdrisBound{n} \IdrisBound{a})} would
have a size quadratic in \IdrisBound{n}.
A smarter representation with perfect sharing would still represent quite an
overhead as observed by Brady, McBride, and McKinna~\cite{DBLP:conf/types/BradyMM03}.

A finite number is a number known to be strictly smaller than a given natural
number. The type \IdrisType{Fin} is indexed by said bound.
Its \IdrisData{Z} constructor models \IdrisData{0} and is bound by any
non-zero bound, and its \IdrisData{S} constructor takes a number bound by
\IdrisBound{n} and returns its successor, bound by
(\IdrisData{1} \IdrisFunction{+} \IdrisBound{n}).
A naïve compilation would here also lead to a runtime representation suffering
from a quadratic blowup.

\ExecuteMetaData[Lookup.idr.tex]{fin}

This leads us to the definition of the \IdrisFunction{lookup} function.
Provided a vector of size \IdrisBound{n} and a finite number \IdrisBound{k} bound
by this same \IdrisBound{n}, we can define a \emph{total} function looking up the
value stored at position \IdrisBound{k} in the vector.
It is guaranteed to return a value.
Note that we do not need to consider the case of the empty vector in the pattern
matching clauses as all of the return types of the \IdrisType{Fin} constructors force
the index to be non-zero and, because the vector and the finite number talk about the
same \IdrisBound{n}, having an empty vector would automatically imply having a value
of type (\IdrisType{Fin} \IdrisData{0}) which is self-evidently impossible.

\ExecuteMetaData[Lookup.idr.tex]{vectlookup}

Thanks to our indexed family, we have gained the ability to define a function that cannot
possibly fail, as well as the ability to only talk about the pattern matching clauses
that make sense.
This seemed to be at the cost of efficiency but luckily for us there has already been
extensive work on erasure to automatically detect redundant
data~\cite{DBLP:conf/types/BradyMM03} or data that will not be used at
runtime~\cite{DBLP:journals/pacmpl/Tejiscak20}.

\subsection{Optimising \IdrisType{Vect}, \IdrisType{Fin}, and \IdrisFunction{lookup}}

An analysis in the style of Brady, McBride, and McKinna's~\cite{DBLP:conf/types/BradyMM03}
can solve the quadratic blowup highlighted above by observing
that the natural number a vector is indexed by is entirely determined by the spine of
the vector. In particular, the length of the tail does not need to be stored
as part of the constructor: it can be reconstructed as the predecessor of the length
of the overall vector. As a consequence, a vector can be adequately represented at
runtime by a pair of a natural number and a list. Similarly a bounded number can be
adequately represented by a pair of natural numbers. Putting all of this together and
remembering that the vector and the finite number share the same \IdrisBound{n},
\IdrisFunction{lookup} can be compiled to a function taking two natural numbers and a list.
In \idris{} we would write the optimised \IdrisFunction{lookup} as follows (we use the
\IdrisKeyword{partial} keyword because this transformed version is not total at that type).

\ExecuteMetaData[Lookup.idr.tex]{erasedvectlookup}

We can see in the second clause that the recursive call is performed on the tail of
the list (formerly vector) and so the first argument to \IdrisFunction{lookup}
corresponding to the vector's size is decreased by one. The invariant, despite not
being explicit anymore, is maintained.

A Tejiščák-style analysis~\cite{DBLP:journals/pacmpl/Tejiscak20} can additionally
notice that the lookup function never makes
use of the bound's value and drop it entirely. This leads to the lookup function on
vectors being compiled to its partial-looking counterpart acting on lists.

\ExecuteMetaData[Lookup.idr.tex]{finallookup}

Even though this is in our opinion a pretty compelling example of erasing away the
apparent complexity introduced by inductive families, this approach has two drawbacks.

Firstly, it relies on the fact that the compiler can and will automatically perform
these optimisations.
But nothing in the type system prevents users from inadvertently using a value they
thought would get erased, thus preventing the Tejiščák-style optimisation from firing.
In performance-critical settings, users may rather want to state their intent
explicitly and be kept to their word by the compiler in exchange for predictable
and guaranteed optimisations.

Secondly, this approach is intrinsically limited to transformations that preserve the
type's overall structure: the runtime data structures are simpler but very similar still.
We cannot expect much better than that.
It is so far unrealistic to expect e.g. a change of representation to use a
balanced binary tree instead of a list in order to get logarithmic lookups rather
than linear ones.

\subsection{No Magic Solution}

Even if we are able to obtain a more compact representation of the inductive
family at runtime through enough erasure, this does not guarantee runtime efficiency.
As the \coq{} manual~\cite{Coq:manual} reminds its users, extraction does not magically
optimise away a user-defined quadratic multiplication algorithm when extracting unary
natural numbers to an efficient machine representation.
In a pragmatic move, \coq{}, \agda{}, and \idris{} all have ad-hoc rules to replace
convenient but inefficiently implemented numeric functions with asymptotically faster
counterparts in the target language.

However this approach is not scalable: if we may be willing to extend our trusted core to a
high quality library for unbounded integers, we do not want to replace
our code only proven correct thanks to complex invariants with a wildly different
untrusted counterpart purely for efficiency reasons.

In this paper we use Quantitative Type
Theory~\cite{DBLP:conf/birthday/McBride16,DBLP:conf/lics/Atkey18}
as implemented in \idris{}~\cite{DBLP:conf/ecoop/Brady21} to bridge the gap between
an invariant-rich but inefficient representation based on an inductive family and
an unsafe but efficient implementation using low-level primitives.
Inductive families allow us to
\emph{view}~\cite{DBLP:conf/popl/Wadler87,DBLP:journals/jfp/McBrideM04}
the runtime relevant information encoded in the low-level and efficient representation
as an information-rich compile time data structure. Moreover the quantity annotations
guarantee that this additional information will be erased away during compilation.

%% file: QTT.tex
\section{Some Key Features of \idris}\label{sec:quantitativeTT}

\idris{} implements Quantitative Type Theory,
a Martin-Löf type theory enriched with a semiring of quantities
classifying the ways in which values may be used.
In a type, each binder is annotated with the quantity by which its
argument must abide.

\subsection{Quantities}

A value may be \emph{runtime irrelevant}, \emph{linear}, or \emph{unrestricted}.

\emph{Runtime irrelevant} values (\IdrisKeyword{0} quantity) cannot possibly influence
control flow as they will be erased entirely during compilation.
This forces the language to impose strong restrictions on pattern-matching over these
values.
Typical examples are types like the \IdrisBound{a} parameter in (\IdrisType{List} \IdrisBound{a}),
or indices like the natural number \IdrisBound{n} in
(\IdrisType{Vect} \IdrisBound{n} \IdrisBound{a}).
These are guaranteed to be erased at compile time. The advantage over a Tejiščák-style
analysis is that users can state their intent that an argument ought to be runtime
irrelevant and the language will insist that it needs to be convinced it indeed is.

\emph{Linear} values (\IdrisKeyword{1} quantity) have to be used exactly once.
Typical examples include the \IdrisData{\%World} token used by \idris{} to implement the
\IdrisType{IO} monad à la Haskell, or file handles that cannot be discarded without first explicitly
closing the file.
At runtime these values can be updated destructively. We will not use linearity in this paper.

Last, \emph{unrestricted} values (denoted by no quantity annotation) can flow into any
position, be duplicated or thrown away.
They are the usual immutable values of functional programming.

The most basic of examples mobilising both the runtime irrelevance and unrestricted
quantities is the identity function.

\ExecuteMetaData[QuantitativeTT.idr.tex]{identity}

Its type starts with a binder using curly braces.
This means it introduces an implicit variable that does not need to be filled in by
the user at call sites and will be reconstructed by unification.
The variable it introduces is named \IdrisBound{a} and
has type \IdrisType{Type}. It has the \IdrisKeyword{0} quantity annotation which means
that this argument is runtime irrelevant and so will be erased during compilation.

The second binder uses parentheses. It introduces an explicit variable whose name
is \IdrisBound{x} and whose type is the type \IdrisBound{a} that was just bound. It has
no quantity annotation which means it will be an unrestricted variable.

Finally the return type is the type \IdrisBound{a} bound earlier. This is, as expected,
a polymorphic function from \IdrisBound{a} to \IdrisBound{a}. It is implemented using
a single clause that binds \IdrisBound{x} on the left-hand side and immediately returns
it on the right-hand side.

If we were to try to annotate the binder for \IdrisBound{x} with a \IdrisKeyword{0}
quantity to make it runtime irrelevant then \idris{} would
rightfully reject the definition.
The following \IdrisKeyword{failing} block shows part of the error message complaining
that \IdrisBound{x} cannot be used at an unrestricted quantity on the right-hand side.

\ExecuteMetaData[QuantitativeTT.idr.tex]{invalididentity}



\subsection{Proof Search}\label{sec:proofsearch}

In \idris{}, Haskell-style ad-hoc polymorphism~\cite{DBLP:conf/popl/WadlerB89}
is superseded by a more general proof search mechanism.
Instead of having blessed notions of type classes, instances and constraints,
the domain of any dependent function type can be marked as \IdrisKeyword{auto}.
This signals to the compiler that the corresponding argument will be an implicit
argument and that it should not be reconstructed by unification alone but rather by
proof search.
The search algorithm will use the appropriate user-declared hints as well as
the local variables in scope.

By default, a datatype's constructors are always added to the database of hints.
And so the following declaration brings into scope both an indexed family
\IdrisType{So} of proofs that a given boolean is \IdrisData{True}, and a unique
constructor \IdrisData{Oh} that is automatically added as a hint.

\ExecuteMetaData[QuantitativeTT.idr.tex]{so}

As a consequence, we can for instance define a record type specifying what it
means for \IdrisBound{n} to be an even number by storing its \IdrisFunction{half}
together with a proof that is both runtime irrelevant and filled in by proof search.
Because (\IdrisData{2} \IdrisFunction{*} \IdrisData{3} \IdrisFunction{==} \IdrisData{6})
computes to \IdrisData{True}, \idris{} is able to fill-in the missing proof in the
definition of \IdrisFunction{even6} using the \IdrisData{Oh} hint.

\noindent
\begin{minipage}[t]{0.55\textwidth}
\ExecuteMetaData[QuantitativeTT.idr.tex]{even}
\end{minipage}\hfill
\begin{minipage}[t]{0.4\textwidth}
\ExecuteMetaData[QuantitativeTT.idr.tex]{four}
\end{minipage}

We will use both \IdrisType{So} and the \IdrisKeyword{auto} mechanism in
\cref{sec:thininginternal}.

\subsection{Application: \IdrisType{Vect}, as \IdrisType{List}}\label{sec:vectaslist}

We can use the features of Quantitative Type Theory to give an implementation
of \IdrisType{Vect} that is guaranteed to erase to a \IdrisType{List} at runtime
independently of the optimisation passes implemented by the compiler.
The advantage over the optimisation passes described in \cref{sec:optimisation-example}
is that the user has control over the runtime representation and does not need to
rely on these optimisations being deployed by the compiler.

The core idea is to make the slogan `a vector is a length-indexed list' a reality
by defining a record packing together the \IdrisFunction{encoding} as a list and
a proof its length is equal to the expected \IdrisType{Nat} index.
This proof is marked as runtime irrelevant to ensure that the list is the only
thing remaining after compilation.

\ExecuteMetaData[VectAsList.idr.tex]{vect}

\paragraph{Smart constructors}
Now that we have defined vectors,
we can recover the usual building blocks for vectors by defining smart
constructors, that is to say functions \IdrisFunction{Nil} and
\IdrisFunction{(::)} that act as replacements for the inductive
family's data constructors.

\ExecuteMetaData[VectAsList.idr.tex]{nil}

The smart constructor \IdrisFunction{Nil} returns an empty vector.
It is, unsurprisingly, encoded as the empty list (\IdrisData{[]}).
Because (\IdrisFunction{length} \IdrisData{[]}) statically computes to
\IdrisData{Z}, the proof that the encoding is valid can be discharged by
reflexivity.

\ExecuteMetaData[VectAsList.idr.tex]{cons}

Using \IdrisFunction{(::)} we can combine a head and a tail of size \IdrisBound{n}
to obtain a vector of size (\IdrisData{S} \IdrisBound{n}).
The encoding is obtained by consing the head in front of the tail's encoding
and the proof this is valid
{(\IdrisFunction{cong} \IdrisData{S} \IdrisBound{eq})}
uses the fact that propositional
equality is a congruence and that
(\IdrisFunction{length} (\IdrisBound{x} \IdrisData{::} \IdrisBound{xs}))
computes to (\IdrisData{S} (\IdrisFunction{length} \IdrisBound{xs})).

\paragraph{View}
Now that we know how to build vectors, we demonstrate that we can also take
them apart using a view.

A view for a type $T$, in the sense of Wadler~\cite{DBLP:conf/popl/Wadler87},
and as refined by McBride and McKinna~\cite{DBLP:journals/jfp/McBrideM04},
is an inductive family $V$ indexed by $T$ together with a total function
mapping every element $t$ of $T$ to a value of type ($V t$).
This simple gadget provides a powerful, user-extensible, generalisation of
pattern-matching.
Patterns are defined inductively as either a pattern variable, a forced term
(i.e. an arbitrary expression that is determined by a constraint arising from
another pattern), or a data constructor fully applied to subpatterns.
In contrast, the return indices of an inductive family's constructors can be
arbitrary expressions.

In the case that interests us, the view allows us to emulate `matching'
on which of the two smart constructors \IdrisFunction{Nil} or \IdrisFunction{(::)}
was used to build the vector being taken apart.

\ExecuteMetaData[VectAsList.idr.tex]{dataview}

The inductive family \IdrisType{View} is indexed by a vector and has two
constructors corresponding to the two smart constructors.
We use \idris{}'s overloading capabilities to give each of the
\IdrisType{View}'s constructors the name of the smart constructor
it corresponds to.
By pattern-matching on a value of type (\IdrisType{View} \IdrisBound{xs}),
we will be able to break \IdrisBound{xs} into its constitutive parts and
either observe it is equal to \IdrisFunction{Nil} or recover its head
and its tail.

\ExecuteMetaData[VectAsList.idr.tex]{view}

The function \IdrisFunction{view} demonstrates that we can always tell which
constructor was used by inspecting the \IdrisFunction{encoding} list. If it
is empty, the vector was built using the \IdrisFunction{Nil} smart constructor.
If it is not then we got our hands on the head and the tail of the encoding
and (modulo some re-wrapping of the tail) they are effectively the head and the
tail that were combined using the smart constructor.

\subsubsection{Application: \IdrisFunction{map}}

We can then use these constructs to implement the function \IdrisFunction{map}
on vectors without ever having to explicitly manipulate the encoding.
The maximally sugared version of \IdrisFunction{map} is as follows:

\ExecuteMetaData[VectAsList.idr.tex]{map}

On the left-hand side the view lets us seamlessly pattern-match on the input
vector.
Using the \IdrisKeyword{with} keyword we have locally modified the function
definition so that it takes an extra argument, here the result of the intermediate
computation (\IdrisFunction{view} \IdrisBound{xs}).
Correspondingly, we have two clauses matching on this extra argument;
the symbol \IdrisKeyword{|} separates the original left-hand side
(here elided using \IdrisKeyword{\KatlaUnderscore{}} because it is exactly the
same as in the parent clause) from the additional pattern.
This pattern can
either have the shape \IdrisData{[]} or (\IdrisBound{hd} \IdrisData{::} \IdrisBound{tl})
and, correspondingly, we learn that \IdrisBound{xs} is either \IdrisFunction{[]} or
(\IdrisBound{hd} \IdrisFunction{::} \IdrisBound{tl}).

On the right-hand side the smart constructors let us build the output vector.
Mapping a function over the empty vector yields the empty vector while mapping
over a cons node yields a cons node whose head and tail have been appropriately
modified.

This sugared version of \IdrisFunction{map} is equivalent to the following more
explicit one:

\ExecuteMetaData[WithExpanded.idr.tex]{map}

In the parent clause we have explicitly bound \IdrisBound{xs}
instead of merely introducing an alias for it by writing
{(\IdrisBound{xs}\IdrisKeyword{@}\IdrisKeyword{\_})}
and so we will need to be explicit about the ways in which this
pattern is refined in the two with-clauses.

In the with-clauses, we have explicitly repeated the refined version
of the parent clause's left-hand side. In particular we have used dotted
patterns to insist that \IdrisBound{xs} is now entirely \emph{forced}
by the match on the result of (\IdrisFunction{view} \IdrisBound{xs}).

We have seen that by matching on the result of the
(\IdrisFunction{view} \IdrisBound{xs}) call,
we get to `match' on \IdrisBound{xs} as if \IdrisType{Vect} were an
inductive type.
This is the power of views.

\subsubsection{Application: \IdrisFunction{lookup}}

The type (\IdrisType{Fin} \IdrisBound{n}) can similarly be represented by a
single natural number and a runtime irrelevant proof that it is bound by
\IdrisBound{n}.
We leave these definitions out, and invite the curious reader
to either attempt to implement them for themselves or look at the accompanying code.

Bringing these definitions together, we can define a \IdrisFunction{lookup}
function which is  similar to the one defined in \cref{sec:optimisation-example}.

\ExecuteMetaData[LookupRefactor.idr.tex]{lookup}

We are seemingly using \IdrisFunction{view} at two different types (\IdrisType{Vect}
and \IdrisType{Fin} respectively) but both occurrences actually refer to separate
functions: \idris{} lets us overload functions and performs type-directed disambiguation.

For pedagogical purposes, this sugared version of \IdrisFunction{lookup} can
also be expanded to a more explicit one that demonstrates the views' power.

\ExecuteMetaData[WithExpanded.idr.tex]{lookup}

The main advantage of this definition is that, based on its type alone, we know
that this function is guaranteed to be processing a list and a single natural
number at runtime.
This efficient runtime representation does not rely on the assumption that state
of the art optimisation passes will be deployed.

We have seen some of \idris{}'s powerful features and how they can be leveraged
to empower users to control the runtime representation of the inductive families
they manipulate.
This simple example only allowed us to reproduce the performance that could already
be achieved by compilers deploying state of the art optimisation passes.
In the following sections, we are going to see how we can use the same core ideas
to compile an inductive family to a drastically different runtime representation
while keeping good high-level ergonomics.

%% file: CdB.tex
\section{Thinnings, cooked two ways}\label{sec:codebruijn}

\iftoggle{BLIND}
{In their work on \typos~\cite{MANUAL:talk/types/Allais22}, a domain specific language
to define concurrent typecheckers and elaborators, the authors go out of their way
to avoid using inductive families because of their inefficient runtime representation.
}
{We experienced a major limitation of compilation of inductive families
during our ongoing development of
\typos~\cite{MANUAL:talk/types/Allais22}, a domain specific language
to define concurrent typecheckers and elaborators.
}
Core to this project is the definition of actors manipulating a generic notion
of syntax with binding.
Internally the terms of this syntax with binding are based on a co-de Bruijn
representation (an encoding we will explain below) which relies heavily on
thinnings.
A thinning (also known as an Order Preserving
Embedding~\cite{MANUAL:phd/nott/Chapman09})
between a source and a target scope is an order preserving injection
of the smaller scope into the larger one.
They are usually represented using an inductive family.
The omnipresence of thinnings in the co-de Bruijn representation makes their
runtime representation a performance critical matter.

Let us first remind the reader of the structure of abstract syntax trees in a
named, a de Bruijn, and a co-de Bruijn representation. We will then discuss two
representations of thinnings: a safe and convenient one as an inductive family,
and an unsafe but efficient encoding as a pair of arbitrary precision integers.

\subsection{Named, de Bruijn, and co-de Bruijn syntaxes}

In this section we will use
the $S$ combinator ($\lambda g. \lambda f. \lambda x. g x (f x)$)
as a running example and represent
terms using a syntax tree whose constructor nodes are circles and variable nodes
are squares.
To depict the $S$ combinator we will only need $\lambda{}$-abstraction and
application (rendered \$) nodes. A constructor's arguments become its children
in the tree.
The tree is laid out left-to-right and a constructor's arguments are displayed
top-to-bottom.

\paragraph{Named syntax}
The first representation is using explicit names. Each binder has an associated
name and each variable node carries a name. A variable refers to the closest enclosing
binder which happens to be using the same name.

\ExecuteMetaData[ast.tex]{named}

To check whether two terms are structurally equivalent (\emph{$\alpha$-equivalence})
potentially requires renaming bound names.
In order to have a simple and cheap $\alpha$-equivalence check we can instead opt
for a nameless representation.

\paragraph{De Bruijn syntax}
An abstract syntax tree based on de Bruijn indices~\cite{MANUAL:journals/math/debruijn72}
replaces names with natural numbers counting the number of binders separating a variable
from its binding site.
The $S$ combinator is now written $(\lambda\, \lambda\, \lambda\, 2\, 0\, (1\, 0))$.

You can see in the following graphical depiction that
$\lambda$-abstractions do not carry a name anymore and that variables are simply pointing
to the binder that introduced them. We have left the squares empty but in practice
the various coloured arrows would be represented by a natural number.
For instance the {\color{magenta}dashed magenta} one corresponds to $1$
because you need to ignore one $\lambda{}$-abstraction
(the {\color{orange}orange} one) on your way towards the root of the tree
before you reach the corresponding magenta binder.

\ExecuteMetaData[ast.tex]{debruijn}

To check whether a subterm does not mention a given set of variables
(a \emph{thickening} test, the opposite of a \emph{thinning} which extends the
current scope with unused variables), you need to traverse the whole term.
In order to have a simple cheap thickening test we can ensure that each subterms
knows precisely what its \emph{support} is and how it embeds
in its parent's.

\paragraph{Co-de Bruijn syntax}
In a co-de Bruijn
representation~\cite{DBLP:journals/corr/abs-1807-04085} each subterm
selects exactly the variables that stay in scope for that term,
and so a variable constructor ultimately refers to the only variable still
in scope by the time it is reached.
This representation ensures that we know precisely what the scope of a given term
currently is.

In the following graphical rendering, we represent thinnings as lists of full
($\bullet$) or empty ($\circ$) discs depending on whether the corresponding
variable is either kept or discarded.
For instance the thinning represented by
$\color{blue}{\circ}\color{magenta}{\bullet}\color{orange}{\bullet}$
throws the {\color{blue}blue} variable away, and keeps both the
{\color{magenta}magenta} and {\color{orange}orange} ones.

\ExecuteMetaData[ast.tex]{codebruijn}

We can see that in such a representation, each node in the tree stores one
thinning per subterm. This will not be tractable unless we have an efficient
representation of thinnings.

\subsection{The Performance Challenges of co-de Bruijn}~\label{sec:thinningsintypos}

Using the co-de Bruijn approach, a term in an arbitrary context is represented
by the pairing of a term in co-de Bruijn syntax with a thinning from its support
into the wider scope.
Having such a precise handle on each term's support allows us to make operations
such as thinning, substitution, unification, or common sub-expression elimination
more efficient.

Thinning a term does not require us to traverse it anymore.
Indeed, embedding a term in a wider context will not change its support
and so we can simply compose the two thinnings while keeping the term the same.

Substitution can avoid traversing subterms that will not
be changed. Indeed, it can now easily detect when the substitution's domain
does not intersect with the subterm's support.

Unification requires performing thickening tests when we want to
solve a metavariable declared in a given context with a terms seemingly living
in a wider one. We once more do not need to traverse the term to perform this
test, and can simply check whether the outer thinning can be thickened.

Common sub-expression elimination requires us to identify alpha-equivalent terms
potentially living in different contexts. Using a de Bruijn representation, these
can be syntactically different: a variable represented by the natural number $v$
in $\Gamma$ would be $(1+v)$ in $\Gamma,\sigma$ but $(2+v)$ in $\Gamma,\tau,\nu$.
A co-de Bruijn representation, by discarding all the variables not in the support,
guarantees that we can once more use syntactic equality to detect alpha-equivalence.
This encoding is used for instance (albeit unknowingly) by Maziarz, Ellis,
Lawrence, Fitzgibbon, and Peyton-Jones in their
`Hashing modulo alpha-equivalence' work~\cite{DBLP:conf/pldi/MaziarzELFJ21}.

For all of these reasons
\iftoggle{BLIND}
{the \typos~\cite{MANUAL:talk/types/Allais22} authors have, as we mentioned earlier,
opted for a co-de Bruijn representation.

}{
we have, as we mentioned earlier, opted for a co-de Bruijn
representation in the implementation of \typos~\cite{MANUAL:talk/types/Allais22}.
}
And so it is crucial for performance that
\iftoggle{BLIND}{they}{we} have a compact representation of thinnings.

\subsubsection{Thinnings in \typos}

\iftoggle{BLIND}{The authors}{We}
first carefully worked out the trickier parts of the implementation in Agda before
porting the resulting code to Haskell.
This process \iftoggle{BLIND}{highlights}{highlighted}
a glaring gap between on the one hand the experiments done
using a strongly typed inductive representation of thinnings and on the other hand
their more efficient but unsafe encoding in Haskell.

\paragraph{Agda}
The Agda-based experiments use inductive families that make the key invariants
explicit which helps tracking complex constraints and catches design flaws at
typechecking time.
The indices guarantee that we always transform the thinnings appropriately when
we add or remove bound variables. In \idris{}, the inductive family representation
of thinnings would be written:

\ExecuteMetaData[Thinnings.idr.tex]{family}
The \IdrisType{Thinning} family is indexed by two scopes (represented as snoclists
i.e. lists that are extended from the right, just like contexts in inference rules):
\IdrisBound{sx} the tighter scope and \IdrisBound{sy} the wider one.
The \IdrisData{Done} constructor corresponds to a thinning from the empty scope to
itself (\IdrisData{[<]} is \idris{} syntactic sugar for the empty snoclist),
and \IdrisData{Keep} and \IdrisData{Drop} respectively extend a given thinning
by keeping or dropping the most local variable (\IdrisData{:<} is the `snoc'
constructor, a sort of flipped `cons').
The `name' (\IdrisBound{x} of type \IdrisBound{a}) is marked with the quantity
\IdrisKeyword{0} to ensure it is erased at compile time (cf. \cref{sec:quantitativeTT}).

During compilation, \idris{} would erase the families' indices as they are forced
(in the sense of Brady, McBride, and McKinna~\cite{DBLP:conf/types/BradyMM03}),
and drop the constructor arguments marked as runtime irrelevant.
The resulting inductive type would be the following simple data type.

\ExecuteMetaData[Thinnings.idr.tex]{bare}

At runtime this representation is therefore essentially a linked list of booleans
(\IdrisData{Done} being \IdrisData{Nil}, and \IdrisData{Keep} and \IdrisData{Drop}
respectively (\IdrisData{True} \IdrisData{::}) and (\IdrisData{False} \IdrisData{::})).

\paragraph{Haskell}
The Haskell implementation uses this observation and picks a packed encoding
of this list of booleans as a pair of integers.
One integer represents the length \IdrisBound{n} of the list, and the other
integer's \IdrisBound{n} least significant bits encode the list as a bit pattern
where \IdrisData{1} is \IdrisData{Keep} and \IdrisData{0} is \IdrisData{Drop}.

Basic operations on thinnings are implemented by explicitly manipulating individual bits.
It is not indexed and thus all the invariant tracking has to be done by hand.
This has led to numerous and hard to diagnose bugs.

\subsubsection{Thinnings in \idris}

\idris{} is a self-hosting language whose core datatype is currently based on
a well-scoped de Bruijn representation.
This precise indexing of terms by their scope helped entirely eliminate a whole
class of bugs that plagued Idris 1's unification machinery.

If \iftoggle{BLIND}{\idris{}}{we} were to switch to a co-de Bruijn
representation for \iftoggle{BLIND}{the}{our} core language
\iftoggle{BLIND}{the core developers would want}{we would want},
and should be able, to have the best of both worlds:
a safe \emph{and} efficient representation!

Thankfully \idris{} implements Quantitative Type Theory (QTT) which gives us a
lot of control over what is to be runtime relevant and what is to be erased
during compilation.
This should allow us to insist on having a high-level interface that resembles
an inductive family while ensuring that everything but a pair of integers is erased
at compile time.
We will exploit the key features of QTT presented in \cref{sec:quantitativeTT}
to have our cake and eat it.

%% file: efficient.tex
\section{An Efficient Invariant-Rich Representation}\label{sec:efficient}

We can combine both approaches highlighted in \cref{sec:thinningsintypos}
by defining a record parameterised by a source
(\IdrisBound{sx}) and target (\IdrisBound{sy}) scopes corresponding to the two
ends of the thinnings, just like we would for the inductive family. This record
packs two numbers and a runtime irrelevant proof.

Firstly, we have a natural number called \IdrisFunction{bigEnd} corresponding
to the size of the big end of the thinning (\IdrisBound{sy}).
We are happy to use a (unary) natural number here because we know that \idris{}
will compile it to an unbounded integer.

Secondly, we have an integer called \IdrisFunction{encoding} corresponding to
the thinning represented as a bit vector stating, for each variable, whether
it is kept or dropped. We only care about the integer's \IdrisFunction{bigEnd}
least significant bits and assume the rest is set to 0.

Thirdly, we have a runtime irrelevant proof \IdrisFunction{invariant} that
\IdrisFunction{encoding} is indeed a valid encoding of size \IdrisFunction{bigEnd}
of a thinning from \IdrisBound{sx} to \IdrisBound{sy}. We will explore the
definition of the relation \IdrisType{Invariant} later on
in \cref{sec:thininginternal}.

\ExecuteMetaData[Thin.idr.tex]{thin}

The first sign that this definition is adequate is our ability to construct
any valid thinning. We demonstrate it is the case by introducing functions
that act as smart constructor analogues for the inductive family's data
constructors.

\subsection{Smart Constructors for \IdrisType{Th}}

The first and simplest one is \IdrisFunction{done}, a function that packs a pair of
\IdrisData{0} (the size of the big end, and the empty encoding) together with a proof
that it is an adequate encoding of the thinning from the empty scope to itself.
In this instance, the proof is simply the \IdrisData{Done} constructor.

\ExecuteMetaData[Thin.idr.tex]{done}

To implement both \IdrisFunction{keep} and \IdrisFunction{drop}, we are going to need
to perform bit-level manipulations.
These are made easy by \idris{}'s \IdrisType{Bits} interface which provides us with
functions to
shift the bit patterns left or right (\IdrisFunction{shiftl}, \IdrisFunction{shiftr}),
set or clear bits at specified positions (\IdrisFunction{setBit}, \IdrisFunction{clearBit}),
take bitwise logical operations like disjunction (\IdrisFunction{.|.})
or conjunction (\IdrisFunction{.\&.}), etc.

In both  \IdrisFunction{keep} and \IdrisFunction{drop}, we need to extend the
encoding with an additional bit.
For this purpose we introduce the \IdrisFunction{cons} function which takes a bit
$b$ and an existing encoding $bs$ and returns the new encoding $\bitcons{bs}{b}$.

\ExecuteMetaData[Data/Bits/Integer.idr.tex]{cons}

No matter what the value of the new bit is, we start by shifting the encoding to
the left to make space for it; this gives us \IdrisBound{bs0} which contains the
bit pattern $\bitcons{bs}{0}$.
If the bit is \IdrisData{True} then we need to additionally set the bit at position
$0$ to obtain $\bitcons{bs}{1}$. Otherwise if the bit is \IdrisData{False}, we can readily
return the $\bitcons{bs}{0}$ encoding obtained by left shifting.
The correctness of this function is backed by two lemma:
testing the bit at index $0$ after consing amounts to returning the cons'd bit,
and shifting the cons'd encoding to the right takes us back to the unextended
encoding.

\ExecuteMetaData[Data/Bits/Integer.idr.tex]{testBit0Cons}
\ExecuteMetaData[Data/Bits/Integer.idr.tex]{consShiftR}

The \IdrisFunction{keep} smart constructor demonstrates that from a thinning from
\IdrisBound{sx} to \IdrisBound{sy} and a runtime irrelevant variable \IdrisBound{x}
we can compute a thinning from the extended source scope
(\IdrisBound{sx} \IdrisData{:<} \IdrisBound{x}) to the target scope
(\IdrisBound{sy} \IdrisData{:<} \IdrisBound{x}) where \IdrisBound{x} was kept.

\ExecuteMetaData[Thin.idr.tex]{keep}

The outer scope has grown by one variable and so we increment \IdrisFunction{bigEnd}.
The encoding is obtained by \IdrisFunction{cons}-ing the boolean \IdrisData{True} to
record the fact that this new variable is kept.
Finally, we use the two lemmas shown above to convince \idris{} the invariant
has been maintained.

Similarly the \IdrisFunction{drop} function demonstrates that we can compute a
thinning getting rid of the variable \IdrisBound{x} freshly added to the target
scope.

\ExecuteMetaData[Thin.idr.tex]{drop}

We once again increment the \IdrisFunction{bigEnd}, use \IdrisFunction{cons} to
record that the variable is being discarded and use the lemmas ensuring its
correctness to convince \idris{} the invariant is maintained.

We can already deploy these smart constructors to implement functions producing
thinnings. We use \IdrisFunction{which} as our example. It is a filter-like
function that returns a dependent pair
containing the elements that satisfy a boolean predicate
together with a proof that there is a thinning embedding
them back into the input snoclist.

\ExecuteMetaData[Thin.idr.tex]{which}

If the input snoclist is empty then the output shall also be, and
\IdrisFunction{done} builds a thinning from \IdrisData{[<]} to itself.
If it is not empty we can perform a recursive call on the tail of the snoclist
and then depending on whether the predicates holds true of the head we can either
\IdrisFunction{keep} or \IdrisFunction{drop} it.

We are now equipped with these smart constructors that allow us to seamlessly
build thinnings.
To recover the full expressive power of the inductive family, we also need to
be able to take these thinnings apart. We are now going to tackle this issue.

\subsection{Pattern Matching on \IdrisType{Th}}

The \IdrisType{View} family is a sum type indexed by a thinning. It has one
data constructor associated to each smart constructor and storing its arguments.

\ExecuteMetaData[Thin.idr.tex]{view}

The accompanying \IdrisFunction{view} function witnesses the fact that any
thinning arises as one of these three cases.

\ExecuteMetaData[Thin.idr.tex]{viewtotal}

We show the implementation of \IdrisFunction{view} in its entirety but leave
out the technical auxiliary lemma it invokes.
The interested reader can find them in the accompanying material.
We will however inspect the code \IdrisFunction{view} compiles to after erasure
in \cref{sec:compiledview} to confirm that these auxiliary definitions do not
incur any additional runtime cost.

We first start by pattern matching on the \IdrisFunction{bigEnd} of the thinning.
If it is \IdrisData{0} then we know the thinning has to be the empty thinning.
Thanks to an inversion lemma called \IdrisFunction{isDone}, we can collect a lot
of equality proofs:
the encoding \IdrisBound{bs} \emph{has to} be \IdrisData{0},
the source and target scopes \IdrisBound{sx} and \IdrisBound{sy} \emph{have to}
be the empty snoclists,
and the proof \IdrisBound{prf} of the invariant has to be of a specific shape.
Rewriting by these equalities changes the goal type enough
for the typechecker to ultimately see
that the thinning was constructed using the \IdrisFunction{done} smart
constructor and so we can use the view's \IdrisData{Done} constructor.

\ExecuteMetaData[Thin.idr.tex]{viewDone}

In case the thinning is non-empty, we need to inspect the 0-th bit of the encoding
to know whether it keeps or discards its most local variable.
This is done by calling the \IdrisFunction{choose} function which takes a boolean
\IdrisFunction{b} and returns a value of type
{(\IdrisType{Either} (\IdrisType{So} \IdrisBound{b}) (\IdrisType{So} (\IdrisFunction{not} \IdrisBound{b}))}
i.e. we not only inspect the boolean but also record which value we got in a proof
using the \IdrisType{So} family introduced in~\cref{sec:quantitativeTT}.

\ExecuteMetaData[Thin.idr.tex]{viewKeepDrop}

If the bit is set then we know the variable is kept.
And so we can invoke an inversion lemma that will once again provide us with
a lot of equalities that we immediately deploy to reshape the goal's type.
This ultimately lets us assemble a sub-thinning and use the view's
\IdrisData{Keep} constructor.

\ExecuteMetaData[Thin.idr.tex]{viewKeep}

If the bit is not set then we learn that the thinning was constructed using
\IdrisFunction{drop}. We can once again use an inversion lemma to rearrange
the goal and finally invoke the view's \IdrisData{Drop} constructor.

\ExecuteMetaData[Thin.idr.tex]{viewDrop}

We can readily use this function to implement pattern matching functions taking
a thinning apart. We can for instance define \IdrisFunction{kept}, the function
that counts the number of \IdrisFunction{keep} smart constructors used when
manufacturing the input thinning and returns a proof that this is exactly the
length of the source scope \IdrisBound{sx}.

\ExecuteMetaData[Thin.idr.tex]{kept}

We proceed by calling the \IdrisFunction{view} function on the input thinning
which immediately tells us that we only have three cases to consider.
The \IdrisData{Done} case is easily handled because the branch's refined
types inform us that both \IdrisBound{sx} and \IdrisBound{sy} are the
empty snoclist \IdrisData{[<]} whose length is evidently \IdrisData{0}.
In the \IdrisData{Keep} branch we learn that \IdrisBound{sx} has the shape
(\IdrisBound{\KatlaUnderscore} \IdrisData{:<} \IdrisBound{x}) and so we must return the
successor of whatever the result of the recursive call gives us.
Finally in the \IdrisData{Drop} case, \IdrisBound{sx} is untouched and so a
simple recursive call suffices.
Note that the function is correctly detected as total because the target scope
\IdrisBound{sy} is indeed getting structurally smaller at every single recursive
call.
It is runtime irrelevant but it can still be successfully used as a termination
measure by the compiler.

\subsection{The \IdrisType{Invariant} Relation}\label{sec:thininginternal}

We have shown the user-facing \IdrisType{Th} and have claimed that it is possible
to define smart constructors \IdrisFunction{done}, \IdrisFunction{keep},
and \IdrisFunction{drop}, as well as a \IdrisFunction{view} function.
This should become apparent once we show the actual definition of \IdrisType{Invariant}.

\subsubsection{Definition of \IdrisType{Invariant}}

The relation maintains the invariant between the record's
fields \IdrisFunction{bigEnd} (a \IdrisType{Nat})
and \IdrisFunction{encoding} (an \IdrisType{Integer})
and the index scopes \IdrisBound{sx} and \IdrisBound{sy}.
Its definition can favour ease-of-use of runtime efficiency because we statically
know that all of the \IdrisType{Invariant} proofs will be erased during compilation.

\ExecuteMetaData[Thin/Internal.idr.tex]{thinning}

As always, the \IdrisData{Done} constructor is the simplest.
It states that the thinning of size \IdrisData{Z} and encoded as the bit
pattern \IdrisData{0} is the empty thinning.

The \IdrisData{Keep} constructor guarantees that the thinning of
size (\IdrisData{S} \IdrisBound{i}) and encoding \IdrisBound{bs}
represents an injection
from (\IdrisBound{sx} \IdrisData{:<} \IdrisBound{x})
to (\IdrisBound{sy} \IdrisData{:<} \IdrisBound{x})
provided that the bit at position \IdrisData{Z} of \IdrisBound{bs}
is set, and that the rest of the bit pattern (obtained by a right shift
on \IdrisBound{bs}) is a valid thinning of size \IdrisBound{i} from
\IdrisBound{sx} to \IdrisBound{sy}.

The \IdrisData{Drop} constructor is structured the same way, except that
it insists the bit at position \IdrisData{Z} should \emph{not} be set.

We can readily use this relation to prove that some basic encoding are
valid representations of useful thinnings.

\subsubsection{Examples of \IdrisType{Invariant} proofs}

For instance, we can always define a thinning from the empty scope to
an arbitrary scope \IdrisBound{sy}.

\ExecuteMetaData[Thin.idr.tex]{none}

The \IdrisFunction{encoding} of this thinning is \IdrisData{0} because
every variable is being discarded and its \IdrisFunction{bigEnd} is
the length of the outer scope \IdrisBound{sy}.
The proof that this encoding is valid is provided by
the \IdrisFunction{none} lemma proven below.
We once again use \idris{}'s overloading
to give the same to functions that play similar roles but at
different types.

\ExecuteMetaData[Thin/Internal.idr.tex]{none}

The proof proceeds by induction over the outer scope \IdrisBound{sy}. If it
is empty, we can simply use the constructor for the empty thinning.
Otherwise we can invoke \IdrisData{Drop} on the induction hypothesis.
This all typechecks because (\IdrisFunction{testBit} \IdrisData{0} \IdrisData{Z})
computes to \IdrisData{False} and so the \IdrisBound{nb} proof can be constructed
automatically by \idris{}'s proof search (cf. \cref{sec:proofsearch}),
and (\IdrisData{0} \IdrisFunction{`shiftR`} \IdrisData{1}) evaluates to \IdrisData{0}
which means the induction hypothesis has exactly the right type.

The definition of the identity thinning is a bit more involved.
For a scope of size $n$, we are going to need to generate a bit pattern
consisting of $n$ ones.
We define it in two steps.
First, \IdrisFunction{cofull} defines a bit pattern of $k$ zeros followed by
infinitely many ones by shifting $k$ places to the left a bit pattern of ones only.
Then, we obtain \IdrisFunction{full} by taking the complement of \IdrisFunction{cofull}.

\noindent
\begin{minipage}{.45\textwidth}
\ExecuteMetaData[Data/Bits/Integer.idr.tex]{cofull}
\end{minipage}\hfill
\begin{minipage}{.45\textwidth}
\ExecuteMetaData[Data/Bits/Integer.idr.tex]{full}
\end{minipage}

We can then define the identity thinning for a scope of size $n$ by pairing
(\IdrisFunction{full} \IdrisBound{n}) as the
\IdrisFunction{encoding} and \IdrisBound{n}
as the \IdrisFunction{bigEnd}.

\ExecuteMetaData[Thin.idr.tex]{ones}

The bulk of the work is once again in the eponymous lemma proving that this
encoding is valid.

\ExecuteMetaData[Thin/Internal.idr.tex]{ones}

This proof proceeds once more by induction on the scope.
If the scope is empty then once again the constructor for the empty thinning will do.
In the non-empty case, we first appeal to an auxiliary lemma (not shown here) to
construct a proof \IdrisBound{nb} that the bit at position \IdrisData{Z} for a
non-zero \IdrisFunction{full} integer is known to be \IdrisData{True}.
We then need to use another lemma to cast the induction hypothesis which mentions
(\IdrisFunction{full} (\IdrisFunction{length} \IdrisBound{sx})) so that it may be
used in a position where we expect a proof talking about
(\IdrisFunction{full} (\IdrisFunction{length} (\IdrisBound{sx} \IdrisData{:<} \IdrisBound{x}))
\IdrisFunction{`shiftR`} \IdrisData{1}).

\subsubsection{Properties of the \IdrisType{Invariant} relation}

This relation has a lot of convenient properties.

First, it is proof irrelevant: any two proofs that the same
\IdrisBound{i}, \IdrisBound{bs}, \IdrisBound{sx}, and \IdrisBound{sy} are
related are provably equal.
Consequently, equality on \IdrisType{Th} values amounts to equality of
the \IdrisFunction{bigEnd} and \IdrisFunction{encoding} values. In particular
it is cheap to test whether a given thinning is the empty or the
identity thinning.

Second, it can be inverted~\cite{DBLP:conf/types/CornesT95} knowing only two bits:
whether the natural number is empty and what the value of the bit at position
\IdrisData{Z} of the encoding is.
This is what allowed us to efficiently implement the \IdrisFunction{view} function
by using
these two checks and then inverting the \IdrisType{Invariant} proof to gain access
to the proof that the remainder of the thinning's encoding is valid.
We will see in \cref{sec:compiledview} that this leads to efficient runtime code for the view.

\subsection{Choose Your Own Abstraction Level}

Access to both the high-level \IdrisType{View} and the internal \IdrisType{Invariant}
relation means that programmers can pick the level of abstraction at which they
want to work.
They may need to explicitly manipulate bits to implement key operators that are
used in performance-critical paths but can also stay at the highest level for
more negligible operations, or when proving runtime irrelevant properties.

In the previous section we saw simple examples of these bit manipulations when
defining \IdrisFunction{none} (using the constant 0 bit pattern) and
\IdrisFunction{ones} using bit shifting and complement to form an initial segment
of 1s followed by 0s.

Other natural examples include the \emph{meet} and \emph{join} of two thinnings
sharing the same wider scope.
The join can for instance be thought of either as a function defined by induction
on the first thinning and case analysis on the second, emitting a \IdrisData{Keep}
constructor whenever either of the inputs does.
Or we can observe that the bit pattern in the join is exactly the disjunction of
the inputs' respective bit patterns and prove a lemma about the \IdrisType{Invariant}
relation instead.
This can be visualised as follows. In each column, the meet is a
$\bullet$ whenever either of the inputs is.

\[
\begin{array}{c@{~}l}
& \color{blue}{\circ}\color{orange}{\circ}\color{magenta}{\bullet}\color{teal}{\bullet}\color{lightgray}{\circ} \\
  \vee & \color{blue}{\bullet}\color{orange}{\circ}\color{magenta}{\circ}\color{teal}{\bullet}\color{lightgray}{\bullet} \\
  \hline
  & \color{blue}{\bullet}\color{orange}{\circ}\color{magenta}{\bullet}\color{teal}{\bullet}\color{lightgray}{\bullet} \\
\end{array}
\]

The join is of particular importance because it appears when we convert an `opened'
view of a term into its co-de Bruijn counterpart.
As we mentioned earlier, co-de Bruijn terms in an arbitrary scope are represented by
the pairing of a term indexed by its precise support with a thinning embedding this
support back into the wider scope.
When working with such a representation, it is convenient to have access to an
`opened' view where the outer thinning has been pushed inside therefore exposing
the term's top-level constructor, ready for case-analysis.

The following diagram shows the correspondence between an `opened' application node
using the view (the diamond `\$' node) with two subterms both living in the outer scope
and its co-de Bruijn form (the circular `\$' node) with an outer thinning selecting the
term support.

\noindent
\begin{minipage}{.45\textwidth}\center
  \ExecuteMetaData[ast.tex]{opened}
\end{minipage}\hfill
\begin{minipage}{.45\textwidth}\center
  \ExecuteMetaData[ast.tex]{opening}
\end{minipage}

The outer thinning of the co-de Bruijn term is obtained precisely by
computing the join of the respective outer thinnings of the `opened'
application's function and argument.

These explicit bit manipulations will be preserved during compilation and
thus deliver more efficient code.

\subsection{Compiled Code}\label{sec:compiledview}

The following code block shows the JavaScript code that is produced when compiling the
\IdrisFunction{view} function. We chose to use the JavaScript backend rather than e.g.
the ChezScheme one because it produces fairly readable code.
We have modified the backend to also write comments reminding the reader of the type
of the function being defined and the data constructors the natural number tags
correspond to.
These changes are now available to all in \idris{}'s current development version.

The only manual modifications we have performed are the inlining of a function
corresponding to a \IdrisKeyword{case} block, renaming variables and property names
to make them human-readable, introducing the \texttt{\$tail} definitions to make
lines shorter, and slightly changing the layout.

\input{view}

Readers can see that the compilation process has erased all of the indices
and the proofs
showing that the invariant tying the efficient runtime representation to the
high-level specification is maintained.
A thinning is represented at runtime by a JavaScript object with two properties
corresponding to \IdrisType{Th}'s runtime relevant fields: \IdrisFunction{bigEnd}
and \IdrisFunction{encoding}.
Both are storing a JavaScript \texttt{bigInt} (one corresponding to the
\IdrisType{Nat}, the other to the \IdrisType{Integer}).
For instance the thinning [01101] would be at runtime
\mintinline{javascript}{{ bigEnd: 5n, encoding: 13n }}.

The view proceeds in two steps. First if the \texttt{bigEnd} is \texttt{0n}
then we know the thinning is empty and can immediately return the \IdrisData{Done}
constructor.
Otherwise we know the thinning to be non-empty and so we can compute the big end
of its tail (\texttt{\$predBE}) by subtracting one to the non-zero \texttt{bigEnd}.
We can then inspect the bit at position \texttt{0} to decide whether to return a
\IdrisData{Keep} or a \IdrisData{Drop} constructor. This is performed by using a
bit mask to 0-out all the other bits (\texttt{\$th.bigEnd\&1n}) and checking whether
the result is zero.
If it is not equal to 0 then we emit \IdrisData{Keep} and compute the \texttt{\$tail}
of the thinning by shifting the original encoding to drop the 0th bit. Otherwise we
emit \IdrisData{Drop} and compute the same tail.

By running \IdrisFunction{view} on this [01101] thinning, we would get
back (\IdrisData{Keep} [0110]), that is to say
\mintinline{javascript}{{ tag: 1, val: { bigEnd: 4n, encoding: 6n } }}.

Thanks to \idris{}'s implementation of Quantitative Type Theory we have managed
to manufacture a high level representation that can be manipulated like a classic
inductive family using smart constructors and views without giving up an inch of
control on its runtime representation.

The remaining issues such as the fact that we form the view's constructors only
to immediately take them apart thus creating needless allocations can be tackled
by reusing Wadler's analysis (section 12 of \cite{DBLP:conf/popl/Wadler87}).

%% file: view.tex

\begin{minted}{javascript}
/* Thin.Smart.view : (th : Th sx sy) -> View th */
function Thin_Smart_view($th) {
  switch($th.bigEnd) {
    case 0n: return {h: 0 /* Done */};
    default: {
      const $predBE = ($th.bigEnd-1n);
      const $test = choose(notEq(($th.encoding&1n), 0n)));
      switch($test.tag) {
        case 0: /* Left */ {
          const $tail = $th.encoding>>1n;
          return { tag: 1 /* Keep */
                 , val: {bigEnd: $predBE, encoding: $tail}}; }
        case 1: /* Right */ {
          const $tail = $th.encoding>>1n;
          return { tag: 2 /* Drop */
                 , val: {bigEnd: $predBE, encoding: $tail}}; }
}}}}
\end{minted}

%% file: conclusion.tex

\section{Conclusion}\label{sec:conclusion}

We have seen that inductive families provide programmers with ways to root out bugs
by enforcing strong invariants. Unfortunately these families can get in the way of
producing performant code despite existing optimisation passes erasing redundant
or runtime irrelevant data.
This tension has led us to take advantage of Quantitative Type Theory
in order to design a library
combining the best of both worlds: the strong invariants and ease of use of inductive
families together with the runtime performance of explicit bit manipulations.

\subsection{Related Work}

For historical and ergonomic reasons, idiomatic code in \coq{} tends to center programs
written in a subset of the language quite close to OCaml and then prove properties
about these programs in the runtime irrelevant \texttt{Prop} fragment.
This can lead to awkward encodings when the unrefined inputs force the user to consider
cases which ought to be impossible. Common coping strategies involve relaxing the types
to insert a modicum of partiality e.g. returning an option type or taking an additional
input to be used as the default return value.
This approach completely misses the point of type-driven development.
We benefit a lot from having as much information as possible available during
interactive editing.
This information not only helps tremendously getting the definitions right by
ensuring we always maintain vital invariants thus making invalid states
unrepresentable, it also gives programmers access to type-driven tools and automation.
Thankfully libraries such as Equations~\cite{DBLP:conf/itp/Sozeau10,DBLP:journals/pacmpl/SozeauM19}
can help users write more dependently typed programs, by taking care of the complex
encoding required in \coq{}. A view-based approach similar to ours but using \texttt{Prop}
instead of the zero quantity ought to be possible.
We expect that the views encoded this way in Coq will have an even worse computational
behaviour given that Equations uses a sophisticated elaboration process to encode dependent
pattern-matching into Gallina.
However Coq does benefit from good automation support for unfolding lemmas, inversion
principles, and rewriting by equalities which may compensate for the awkwardness introduced
by the encoding.

Prior work on erasure~\cite{DBLP:journals/pacmpl/Tejiscak20} has the advantage of
offering a fully automated analysis of the code. The main inconvenience is that users
cannot state explicitly that a piece of data ought to be runtime irrelevant and so
they may end up inadvertently using it which would prevent its erasure.
Quantitative Type Theory allows us users to explicitly choose what is and is not
runtime relevant, with the quantity checker keeping us true to our word.
This should ensure that the resulting program has a much more predictable complexity.

A somewhat related idea was explored by Brady, McKinna, and Hammond in the context of
circuit design~\cite{DBLP:conf/sfp/BradyMH07}. In their verification work they index
an efficient representation (natural numbers as a list of bits) by its meaning as a
unary natural number. All the operations are correct by construction as witnessed by
the use of their unary counterparts acting as type-level specifications.
In the end their algorithms still process the inductive family instead of working
directly with binary numbers. This makes sense in their setting where they construct
circuits and so are explicitly manipulating wires carrying bits.
By contrast, in our motivating example we really want to get down to actual (unbounded)
integers rather than linked lists of bits.


\subsection{Limitations and Future Work}

Overall we found this case study using \idris{}, a state of the art language
based on Quantitative Type Theory, very encouraging.
The language implementation is still experimental (see for instance
\cref{appendix:limitations} for some of the bugs we found) but none of
the issues are intrinsic limitations.
We hope to be able to push this line of work further, tackling the following
limitations and exploring more advanced use cases.

\subsubsection{Limitations}

Unfortunately it is only \emph{propositionally} true that
(\IdrisFunction{view} (\IdrisFunction{keep} \IdrisBound{th} \IdrisBound{x}))
computes to (\IdrisData{Keep} \IdrisBound{th} \IdrisBound{x}) (and similarly for
\IdrisFunction{done}/\IdrisData{Done} and \IdrisFunction{drop}/\IdrisData{Drop}).
This means that users may need to manually deploy these lemmas when proving the
properties of functions defined by pattern matching on the result of calling the
\IdrisFunction{view} function.
This annoyance would disappear if we had the ability to extend \idris{}'s reduction rules
with user-proven equations as implemented in Agda and formally studied
by Cockx, Tabareau, and Winterhalter~\cite{DBLP:journals/pacmpl/CockxTW21}.

In this paper's case study, we were able to design the core \IdrisType{Invariant}
relation making the invariants explicit in such a way that it would be provably
proof irrelevant.
This may not always be possible given the type theory currently implemented by
\idris{}. Adding support for a proof-irrelevant sort of propositions (see e.g.
Altenkirch, McBride, and Swierstra's work~\cite{DBLP:conf/plpv/AltenkirchMS07})
could solve this issue once and for all.

The \idris{} standard library thankfully gave us access to a polished pure interface
to explicitly manipulate an integer's bits.
However these built-in operations came with no built-in properties whatsoever.
And so we had to postulate a (minimal) set of axioms (see \cref{appendix:postulated})
and prove a lot of useful corollaries ourselves.
There is even less support for other low-level operations such as reading from
a read-only array, or manipulating pointers.

We also found the use of runtime irrelevance (the \IdrisKeyword{0} quantity)
sometimes somewhat frustrating.
Pattern-matching on a runtime irrelevant value in a runtime relevant context
is currently only possible if it is manifest for the compiler that the value
could only arise using one of the family's constructors.
In non-trivial cases this is unfortunately only merely provable rather than
self-evident.
Consequently we are forced to jump through hoops to appease the quantity
checker, and end up defining complex inversion lemmas to bypass these
limitations.
This could be solved by a mix of improvements to the typechecker and
meta-programming using prior ideas on automating
inversion~\cite{DBLP:conf/types/CornesT95,DBLP:conf/types/McBride96,monin:inria-00489412}.

\subsubsection{Future work}

We are planning to explore more memory-mapped representations equipped with a high
level interface.

We already have experimental results demonstrating that we can use a read-only array
as a runtime representation of a binary search tree.
Search can be implemented as a proven-correct high level decision procedure that
is seemingly recursively exploring the "tree".
At runtime however, this will effectively execute like a classic search by dichotomy
over the array.

More generally, we expect that a lot of the work on programming on serialised
data done in LoCal~\cite{DBLP:conf/pldi/VollmerKRS0N19} thanks to specific support
from the compiler can be done as-is in a QTT-based programming language.
Indeed, QTT's type system is powerful enough that tracking these invariants can
be done purely in library code.

In the short term, we would like to design a small embedded domain specific language
giving users the ability to more easily build and take apart products and sums
efficiently represented in the style we presented here.
Staging would help here to ensure that the use of the eDSL comes at no runtime cost.
There are plans to add type-enforced staging to Idris 2, thus really making it the
ideal host language for our project.

Our long term plan is to go beyond read-only data and look at imperative programs
proven correct using separation logic and see how much of this after-the-facts
reasoning can be brought back into the types to enable a high-level
correct-by-construction programming style that behaves the same at runtime.

%% file: acknowledgements.tex
\paragraph{Acknowledgements}
We are grateful to
\iftoggle{BLIND}{[list of colleagues left out of the anonymised version]}
{Conor McBride for discussions pertaining to the fine details
of the unsafe encoding used in TypOS, as well as James McKinna,
Fredrik Nordvall Forsberg,
Ohad Kammar,
and Jacques Carette}
for providing helpful comments and suggestions on early
versions of this paper.

%% file: appendix-axioms.tex
\section{Postulated lemmas for the \IdrisType{Bits} interface}\label{appendix:postulated}

It is often more convenient to reason about integers in terms of their bits.
We define the notion of bitwise equality as the pointwise equality according
to the \IdrisFunction{testBit}.

\ExecuteMetaData[Data/Bits/Integer/Postulated.idr.tex]{extensionalEq}

Our first postulate is a sort of extensionality principle stating that bitwise
equality implies propositional equality.

\ExecuteMetaData[Data/Bits/Integer/Postulated.idr.tex]{extensionally}

This gives us a powerful reasoning principle once combined with axioms
explaining the behaviour of various primitives at the bit level.
This is why almost all of the remaining axioms are expressed in terms
of \IdrisFunction{testBit} calls.

\subsection{Logical operations}

Our first batch of axioms relates logical operations on integers to their
boolean counterparts. This is essentially stating that these operations are
bitwise.

\ExecuteMetaData[Data/Bits/Integer/Postulated.idr.tex]{testBitAnd}
\ExecuteMetaData[Data/Bits/Integer/Postulated.idr.tex]{testBitOr}
\ExecuteMetaData[Data/Bits/Integer/Postulated.idr.tex]{testBitComplement}

Together with the extensionality principle mentioned above this already
allows us to prove for instance that the binary operators are commutative
and associative, that the de Morgan laws hold, or that conjunction distributes
over disjunction.

\subsection{Bit Shifting}

The second set of axiom describes the action of left and right shifting on
bit patterns.

A right shift of size \IdrisBound{k} will drop the \IdrisBound{k} least
significant bits; consequently testing the bit \IdrisBound{i} on the
right-shifted integer amounts to testing the bit
(\IdrisBound{k} \IdrisFunction{+} \IdrisBound{i}) on the original integer.

\ExecuteMetaData[Data/Bits/Integer/Postulated.idr.tex]{testBitShiftR}

A left shift will add \IdrisBound{k} new least significant bits initialised
at \IdrisBound{0}; consequently testing a bit \IdrisBound{i} on the left-shifted
integer will either return \IdrisData{False} if \IdrisBound{i} is strictly less than
\IdrisBound{k}, or the bit at position (\IdrisBound{i} \IdrisFunction{-} \IdrisBound{k})
in the original integer.

For simplicity we state these results without mentioning the `strictly less than'
relation, by considering on the one hand the effect of a non-zero left shift,
and on the other the fact that a left-shift by 0 bits is the identity function.

\ExecuteMetaData[Data/Bits/Integer/Postulated.idr.tex]{testBit0ShiftL}
\ExecuteMetaData[Data/Bits/Integer/Postulated.idr.tex]{testBitSShiftL}
\ExecuteMetaData[Data/Bits/Integer/Postulated.idr.tex]{shiftL0}

\subsection{Bit testing}

The last set of axioms specifies what happens when a bit is set.

Testing a bit other than the one that was set amounts to testing it on the
original integer.

\ExecuteMetaData[Data/Bits/Integer/Postulated.idr.tex]{testSetBitOther}

Finally, we have an axiom stating that the integer
(\IdrisFunction{bit} \IdrisBound{i}) (i.e. $2^i$) is non-zero.

\ExecuteMetaData[Data/Bits/Integer/Postulated.idr.tex]{bitNonZero}

%% file: appendix-limitations.tex
\section{Current Limitations of \idris{}}\label{appendix:limitations}

This challenge, suggested by Jacques Carette, highlights some of the current
limitations of \idris{}.

\subsection{Problem statement}

The goal is to use the \IdrisType{Vect} type defined in~\cref{sec:vectaslist}
and define a view that un-does vector-append.
This is a classic exercise in dependently-typed programming, the interesting
question being whether we can implement the function just as seamlessly with
our encoding.

Vector append can easily be defined by induction over the first vector.

\ExecuteMetaData[VectAsList.idr.tex]{append}

If the first vector is empty we can readily return the second vector.
If it is cons-headed, we can return the head and compute the tail by performing
a recursive call.

Equipped with this definition, we can declare the view type which we call
\IdrisType{SplitAt} by analogy with its weakly typed equivalent processing
lists.
It states that a vector \IdrisBound{xs} of length \IdrisBound{p} can be split
at \IdrisBound{m} if
\IdrisBound{p} happens to be
(\IdrisBound{m} \IdrisFunction{+} \IdrisBound{n})
and \IdrisBound{xs} happens to be
(\IdrisBound{pref} \IdrisFunction{++} \IdrisBound{suff})
where \IdrisBound{pref} and \IdrisBound{suff}'s
respective lengths are \IdrisBound{m} and \IdrisBound{n}.

\ExecuteMetaData[VectAsList.idr.tex]{splitAtrel}

The challenge is to define the function proving that a vector of size
(\IdrisBound{m} \IdrisFunction{+} \IdrisBound{n}) can be split at
\IdrisBound{m}.

\subsection{Failing attempts}

The proof will necessarily go by induction on \IdrisBound{m}, followed by
a case analysis on the input vector and a recursive call in the non-zero
case.

Our first failing attempt successfully splits the natural number, calls the
view on the vector \IdrisBound{xs} to take it apart but then fails when
performing the recursive call to \IdrisFunction{splitAt}.

\ExecuteMetaData[VectAsList.idr.tex]{splitAtFail1}

This reveals an issue in \idris{}'s handling of the interplay between
\IdrisKeyword{@}-patterns and quantities: the compiler arbitrarily decided
to make the alias \IdrisBound{tl} runtime irrelevant only to then complain
that \IdrisBound{tl} is not accessible when we want to perform the recursive
call (\IdrisFunction{splitAt} \IdrisBound{m} \IdrisBound{tl})!

In order to work around this limitation, we decided to let go of
\IdrisKeyword{@}-patterns and write the fully explicit clause ourselves,
using dotted patterns to mark the forced expressions.

\ExecuteMetaData[VectAsList.idr.tex]{splitAtFail2}

The left-hand side now typechecks but the case tree builder fails with a
perplexing error.
This reveals a bug in \idris{}'s implementation of elaboration of
pattern-matching functions to case trees.
Instead of ignoring dotted expressions when building the case tree (these
expressions are forced and so the variables they mention will have necessarily
been bound in another pattern), it attempts to use them to drive the case-splitting
strategy.
This is a well-studied problem and should be fixable by referring to
Cockx and Abel's work~\cite{DBLP:journals/jfp/CockxA20}.

\subsection{Working Around \idris{}'s Limitations}

This leads us to our working solution.
Somewhat paradoxically, working around these \idris{} bugs led us to a more
principled solution whereby the pattern-matching step needed to adjust the
result returned by the recursive call is abstracted away in an auxiliary
function whose type clarifies what is happening.

From an \IdrisBound{m} split on \IdrisBound{xs}, we can easily compute an
(\IdrisData{S} \IdrisBound{m}) split on (\IdrisBound{x} \IdrisFunction{::} \IdrisBound{xs})
by cons-ing \IdrisBound{x} on the prefix.

\ExecuteMetaData[VectAsList.idr.tex]{splitAtcons}

In this auxiliary function, \IdrisBound{xs} is clearly runtime irrelevant and
so the case-splitter will not attempt to inspect it, thus generating the correct
case tree.
We are forced to match further on \IdrisBound{pref} (in particular by making
the equality proof \IdrisData{Refl}) so that just enough computation happens
at the type level for the typechecker to see that things do line up.
A proof irrelevant type of propositional equality would have helped us here.

We can put all of these pieces together and finally get our
\IdrisFunction{splitAt} view.

\ExecuteMetaData[VectAsList.idr.tex]{splitAt}

We do want to reiterate that these limitations are not intrinsic limitations of
the approach, there are just flaws in the current experimental implementation of
the \idris{} language and can and should be remedied.

%% file: thin.bbl
\newcommand{\etalchar}[1]{$^{#1}$}
\begin{thebibliography}{AAM{\etalchar{+}}22}

\bibitem[AAM{\etalchar{+}}22]{MANUAL:talk/types/Allais22}
Guillaume Allais, Malin Altenmüller, Conor McBride, Georgi Nakov,
  Fredrik~Nordvall Forsberg, and Craig Roy.
\newblock {TypOS}: An operating system for typechecking actors.
\newblock In {\em 28th International Conference on Types for Proofs and
  Programs, {TYPES} 2022, June 20-25, 2022, Nantes, France}, 2022.

\bibitem[AMS07]{DBLP:conf/plpv/AltenkirchMS07}
Thorsten Altenkirch, Conor McBride, and Wouter Swierstra.
\newblock Observational equality, now!
\newblock In Aaron Stump and Hongwei Xi, editors, {\em Proceedings of the {ACM}
  Workshop Programming Languages meets Program Verification, {PLPV} 2007,
  Freiburg, Germany, October 5, 2007}, pages 57--68. {ACM}, 2007.

\bibitem[Atk18]{DBLP:conf/lics/Atkey18}
Robert Atkey.
\newblock Syntax and semantics of quantitative type theory.
\newblock In Anuj Dawar and Erich Gr{\"{a}}del, editors, {\em Proceedings of
  the 33rd Annual {ACM/IEEE} Symposium on Logic in Computer Science, {LICS}
  2018, Oxford, UK, July 09-12, 2018}, pages 56--65. {ACM}, 2018.

\bibitem[BMH07]{DBLP:conf/sfp/BradyMH07}
Edwin~C. Brady, James McKinna, and Kevin Hammond.
\newblock Constructing correct circuits: Verification of functional aspects of
  hardware specifications with dependent types.
\newblock In Marco~T. Moraz{\'{a}}n, editor, {\em Proceedings of the Eighth
  Symposium on Trends in Functional Programming, {TFP} 2007, New York City, New
  York, USA, April 2-4. 2007}, volume~8 of {\em Trends in Functional
  Programming}, pages 159--176. Intellect, 2007.

\bibitem[BMM03]{DBLP:conf/types/BradyMM03}
Edwin~C. Brady, Conor McBride, and James McKinna.
\newblock Inductive families need not store their indices.
\newblock In Stefano Berardi, Mario Coppo, and Ferruccio Damiani, editors, {\em
  Types for Proofs and Programs, International Workshop, {TYPES} 2003, Torino,
  Italy, April 30 - May 4, 2003, Revised Selected Papers}, volume 3085 of {\em
  Lecture Notes in Computer Science}, pages 115--129. Springer, 2003.

\bibitem[Bra21]{DBLP:conf/ecoop/Brady21}
Edwin~C. Brady.
\newblock Idris 2: Quantitative type theory in practice.
\newblock In Anders M{\o}ller and Manu Sridharan, editors, {\em 35th European
  Conference on Object-Oriented Programming, {ECOOP} 2021, July 11-17, 2021,
  Aarhus, Denmark (Virtual Conference)}, volume 194 of {\em LIPIcs}, pages
  9:1--9:26. Schloss Dagstuhl - Leibniz-Zentrum f{\"{u}}r Informatik, 2021.

\bibitem[CA20]{DBLP:journals/jfp/CockxA20}
Jesper Cockx and Andreas Abel.
\newblock Elaborating dependent (co)pattern matching: No pattern left behind.
\newblock {\em J. Funct. Program.}, 30:e2, 2020.

\bibitem[CDT22]{Coq:manual}
The {Coq} {Development}~{Team}.
\newblock {\em The {Coq} Proof Assistant Reference Manual, version 8.15.2}, May
  2022.

\bibitem[Cha09]{MANUAL:phd/nott/Chapman09}
James~Maitland Chapman.
\newblock {\em Type checking and normalisation}.
\newblock PhD thesis, University of Nottingham, July 2009.

\bibitem[CT95]{DBLP:conf/types/CornesT95}
Cristina Cornes and Delphine Terrasse.
\newblock Automating inversion of inductive predicates in coq.
\newblock In Stefano Berardi and Mario Coppo, editors, {\em Types for Proofs
  and Programs, International Workshop TYPES'95, Torino, Italy, June 5-8, 1995,
  Selected Papers}, volume 1158 of {\em Lecture Notes in Computer Science},
  pages 85--104. Springer, 1995.

\bibitem[CTW21]{DBLP:journals/pacmpl/CockxTW21}
Jesper Cockx, Nicolas Tabareau, and Th{\'{e}}o Winterhalter.
\newblock The taming of the rew: a type theory with computational assumptions.
\newblock {\em Proc. {ACM} Program. Lang.}, 5({POPL}):1--29, 2021.

\bibitem[dB72]{MANUAL:journals/math/debruijn72}
Nicolaas~Govert de~Bruijn.
\newblock Lambda calculus notation with nameless dummies, a tool for automatic
  formula manipulation, with application to the {C}hurch-{R}osser theorem.
\newblock {\em Indagationes Mathematicae (Proceedings)}, 75(5):381--392, 1972.

\bibitem[Dyb94]{DBLP:journals/fac/Dybjer94}
Peter Dybjer.
\newblock Inductive families.
\newblock {\em Formal Aspects Comput.}, 6(4):440--465, 1994.

\bibitem[McB96]{DBLP:conf/types/McBride96}
Conor McBride.
\newblock Inverting inductively defined relations in {LEGO}.
\newblock In Eduardo Gim{\'{e}}nez and Christine Paulin{-}Mohring, editors,
  {\em Types for Proofs and Programs, International Workshop TYPES'96, Aussois,
  France, December 15-19, 1996, Selected Papers}, volume 1512 of {\em Lecture
  Notes in Computer Science}, pages 236--253. Springer, 1996.

\bibitem[McB16]{DBLP:conf/birthday/McBride16}
Conor McBride.
\newblock I got plenty o' nuttin'.
\newblock In Sam Lindley, Conor McBride, Philip~W. Trinder, and Donald
  Sannella, editors, {\em A List of Successes That Can Change the World -
  Essays Dedicated to Philip Wadler on the Occasion of His 60th Birthday},
  volume 9600 of {\em Lecture Notes in Computer Science}, pages 207--233.
  Springer, 2016.

\bibitem[McB18]{DBLP:journals/corr/abs-1807-04085}
Conor McBride.
\newblock Everybody's got to be somewhere.
\newblock In Robert Atkey and Sam Lindley, editors, {\em Proceedings of the 7th
  Workshop on Mathematically Structured Functional Programming, MSFP@FSCD 2018,
  Oxford, UK, 8th July 2018}, volume 275 of {\em {EPTCS}}, pages 53--69, 2018.

\bibitem[MEL{\etalchar{+}}21]{DBLP:conf/pldi/MaziarzELFJ21}
Krzysztof Maziarz, Tom Ellis, Alan Lawrence, Andrew~W. Fitzgibbon, and
  Simon~Peyton Jones.
\newblock Hashing modulo alpha-equivalence.
\newblock In Stephen~N. Freund and Eran Yahav, editors, {\em {PLDI} '21: 42nd
  {ACM} {SIGPLAN} International Conference on Programming Language Design and
  Implementation, Virtual Event, Canada, June 20-25, 2021}, pages 960--973.
  {ACM}, 2021.

\bibitem[MM04]{DBLP:journals/jfp/McBrideM04}
Conor McBride and James McKinna.
\newblock The view from the left.
\newblock {\em J. Funct. Program.}, 14(1):69--111, 2004.

\bibitem[Mon10]{monin:inria-00489412}
Jean-Fran{\c c}ois Monin.
\newblock {Proof Trick: Small Inversions}.
\newblock In Yves Bertot, editor, {\em {Second Coq Workshop}}, Edinburgh,
  United Kingdom, July 2010. {Yves Bertot}.

\bibitem[SM19]{DBLP:journals/pacmpl/SozeauM19}
Matthieu Sozeau and Cyprien Mangin.
\newblock Equations reloaded: high-level dependently-typed functional
  programming and proving in coq.
\newblock {\em Proc. {ACM} Program. Lang.}, 3({ICFP}):86:1--86:29, 2019.

\bibitem[Soz10]{DBLP:conf/itp/Sozeau10}
Matthieu Sozeau.
\newblock Equations: {A} dependent pattern-matching compiler.
\newblock In Matt Kaufmann and Lawrence~C. Paulson, editors, {\em Interactive
  Theorem Proving, First International Conference, {ITP} 2010, Edinburgh, UK,
  July 11-14, 2010. Proceedings}, volume 6172 of {\em Lecture Notes in Computer
  Science}, pages 419--434. Springer, 2010.

\bibitem[Tej20]{DBLP:journals/pacmpl/Tejiscak20}
Matúš Tejiščák.
\newblock A dependently typed calculus with pattern matching and erasure
  inference.
\newblock {\em Proc. {ACM} Program. Lang.}, 4({ICFP}):91:1--91:29, 2020.

\bibitem[VKR{\etalchar{+}}19]{DBLP:conf/pldi/VollmerKRS0N19}
Michael Vollmer, Chaitanya Koparkar, Mike Rainey, Laith Sakka, Milind Kulkarni,
  and Ryan~R. Newton.
\newblock Local: a language for programs operating on serialized data.
\newblock In Kathryn~S. McKinley and Kathleen Fisher, editors, {\em Proceedings
  of the 40th {ACM} {SIGPLAN} Conference on Programming Language Design and
  Implementation, {PLDI} 2019, Phoenix, AZ, USA, June 22-26, 2019}, pages
  48--62. {ACM}, 2019.

\bibitem[Wad87]{DBLP:conf/popl/Wadler87}
Philip Wadler.
\newblock Views: {A} way for pattern matching to cohabit with data abstraction.
\newblock In {\em Conference Record of the Fourteenth Annual {ACM} Symposium on
  Principles of Programming Languages, Munich, Germany, January 21-23, 1987},
  pages 307--313. {ACM} Press, 1987.

\bibitem[WB89]{DBLP:conf/popl/WadlerB89}
Philip Wadler and Stephen Blott.
\newblock How to make ad-hoc polymorphism less ad-hoc.
\newblock In {\em Conference Record of the Sixteenth Annual {ACM} Symposium on
  Principles of Programming Languages, Austin, Texas, USA, January 11-13,
  1989}, pages 60--76. {ACM} Press, 1989.

\end{thebibliography}
